\documentclass{prop2015}
\usepackage[english]{babel}

\usepackage{graphicx}
\usepackage{amsmath}
\usepackage{hyperref}

\title{Spin-orbit scattering in superconducting nanoparticles}
\author[Y. Alhassid]{Y. Alhassid\inst{1,}\footnote{Corresponding author\quad E-mail:~\textsf{yoram.alhassid@yale.edu}}}
\author[K.\,N. Nesterov]{K. N. Nesterov\inst{2}}
\address[1]{Center for Theoretical Physics, Sloane Physics Laboratory,
Yale University, New Haven, Connecticut 06520, USA}
\address[2]{Department of Physics, University of Wisconsin-Madison, Madison, Wisconsin 53706, USA}
\begin{abstract}
We review interaction effects in chaotic metallic nanoparticles.  Their single-particle Hamiltonian is described by the proper random-matrix ensemble while the dominant interaction terms are invariants under a change of the single-particle basis. In the absence of spin-orbit scattering, the non-trivial  invariants consist of a pairing interaction, which leads to superconductivity in the bulk, and a ferromagnetic exchange interaction.  Spin-orbit scattering breaks spin-rotation invariance and when it is sufficiently strong, the only dominant nontrivial interaction is the pairing interaction. We discuss how the magnetic response of discrete energy levels of the nanoparticle (which can be measured in single-electron tunneling spectroscopy experiments) is affected by such pairing correlations and how it can provide a signature of pairing correlations. We also consider the spin susceptibility of the nanoparticle and discuss how spin-orbit scattering changes the signatures of pairing 
correlations in this observable.  
\end{abstract}

\begin{document}


\maketitle

\section{Introduction}

\begin{figure}
 \includegraphics[width=\columnwidth]{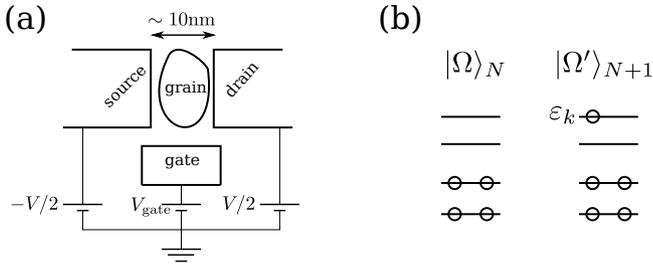}
 \caption{(a) A schematic setup of a single-electron tunneling spectroscopy experiment. A nanoparticle is connected via tunnel barriers to the source and drain electrodes and can also be coupled capacitively to a gate electrode.  (b) An example of an initial and final states of a grain in the absence of electron interactions beyond the classical charging energy. }\label{Fig-intro}
\end{figure}

Ultrasmall metallic grains -- particles whose linear size is only several nanometers -- attracted considerable attention in the mid-1990s, when it became possible to measure their discrete energy levels~\cite{Ralph1995,Black1996,Ralph1997}. Such measurements are carried out using single-electron tunneling spectroscopy techniques~\cite{vonDelft2001}, in which a single-electron transistor is formed by connecting a grain to two metallic leads [see Fig.~\ref{Fig-intro}(a)]. The measured differential conductance displays a series of peaks as a function of the bias voltage, corresponding to opening of new channels for electron tunneling through the nanoparticle. Each channel is associated with a transition of the grain between two many-body eigenstates $|\Omega\rangle_N$ and $|\Omega'\rangle_{N'}$ with particle numbers $N$ and $N'$ differing by one ($|N'-N| = 1$). The positions of these peaks are given by the difference between two many-body energies of these states
\begin{equation}\label{tunneling_spectr_energy}
 \Delta E_{\Omega\Omega'}^{NN`} = E_{\Omega'}^{N'} - E_{\Omega}^N\,.
\end{equation}

In the simplest scenario,  electron-electron interaction in the grain with $N$ electrons is described by the classical charing energy $e^2 N^2/ 2C$, where $e$ is the electron's charge and $C$ is the capacitance of the grain. In this model, each eigenstate is described by occupation numbers of the single-particle orbitals, and  the differential-conductance peaks correspond to tunneling of an electron into or out of a specific orbital [see Fig.~\ref{Fig-intro}(b)]. The measured values (\ref{tunneling_spectr_energy}) reduce then to single-particle energies up to an $N$-dependent constant.

In general, the measured energies (\ref{tunneling_spectr_energy}) do not reduce to single-particle quantities and can potentially provide detailed information about electron interactions in the grain. A particularly interesting case is a superconducting nanoparticle, i.e., a grain made of a material that is superconducting in the bulk. Pairing correlations in such superconducting grains are characterized by the ratio $\Delta/\delta$, where $\Delta$ is the bulk pairing gap and $\delta$ is the mean single-particle level spacing.  The fate of pairing correlations in the crossover from $\Delta/\delta \gg 1$ to $\Delta/\delta \ll 1$ is one of the most interesting questions in studies of superconducting nanoparticles.  When $\Delta/\delta \gg 1$, we are in the limit of the mean-field Bardeen-Cooper-Schrieffer (BCS) theory~\cite{Bardeen1957}, and the spectroscopy measurements in larger grains with $\Delta/\delta >1$ show a gap in the excitation spectrum for even electron number~\cite{Black1996}. When $\Delta/\delta 
< 1$, we are in the so-called fluctuation-dominated regime, where BCS theory  breaks down, and Anderson's criterion~\cite{Anderson1959_jpcs} states that no superconductivity is possible in this regime. Indeed, spectroscopy measurements do not find a gap in this limit~\cite{Black1996}. However,  other  signatures of pairing correlations, e.g., in thermodynamic observables of the grain~\cite{DiLorenzo2000,Falci2000,VanHoucke2006,Nesterov2013}, are predicted to exist even when $\Delta < \delta$. 

In this brief review we focus on superconducting grains with spin-orbit scattering.  In Sec.~\ref{Sec-model} we discuss theoretical models of a chaotic nanoparticle. Its single-electron spectrum fluctuates and follows random-matrix theory (RMT)~\cite{Mehta1991,Beenakker1997,Guhr1998,Alhassid2000}, while the dominant electron interactions in the absence of spin-orbit scattering consist of three non-fluctuating terms: classical charging energy, superconducting pairing term, and a ferromagnetic spin-exchange term. This is described by the so-called universal Hamiltonian~\cite{Kurland2000,Aleiner2002}. In the presence of spin-orbit scattering, spin-rotation symmetry is broken, and, in the limit of strong spin-orbit scattering, the exchange term is absent from the universal Hamiltonian while the pairing interaction survives. In Sec.~\ref{Sec-response}, we review the magnetic response of energies measured in the single-electron tunneling spectroscopy experiments. In the presence of spin-orbit scattering, this 
response is nontrivial: it exhibits level-to-level fluctuations and depends nonlinearly on the magnetic field. We explain how these effects change in the presence of superconducting correlations. In particular, the linear part of the response turns out to be unaffected by pairing, while the quadratic part is highly sensitive to pairing and can be used to probe pairing correlations even in the fluctuation-dominated regime $\Delta<\delta$.  In Sec.~\ref{Sec-thermodyn}, we discuss the spin susceptibility of a superconducting nanoparticle. We review the signatures of pairing correlations in this observable and discuss how they get suppressed by finite spin-orbit scattering.  We conclude in Sec.~\ref{Sec-conclusions}. 

\section{Model}\label{Sec-model}

\subsection{Single-particle Hamiltonian}

The single-particle spectrum and wave functions of a nanoparticle are highly sensitive to atomic-scale irregularities and thus fluctuate. In a grain with chaotic single-particle dynamics or weak disorder, these statistical fluctuations are well described by RMT~\cite{Mehta1991, Beenakker1997,Guhr1998,Alhassid2000}.  RMT is applicable in an energy window of the Thouless energy $E_{\rm Th}$, which is determined by the time it takes for an electron to travel across the nanoparticle. For typical grains used in spectroscopy experiments, $E_{\rm Th} \gg \delta$~\cite{Davidovic1999}, so 
there is a sufficient number of levels for RMT to be meaningful. 

The specific RMT ensemble that describes the statistics of the single-electron Hamiltonian depends on the underlying space-time symmetries~\cite{Mehta1991}. In the absence of spin-orbit coupling and without an external magnetic field, time-reversal symmetry is preserved and the proper RMT ensemble is the Gaussian orthogonal ensemble (GOE). In the presence of spin-orbit scattering, time-reversal continues to be a good symmetry but spin-rotation symmetry is broken, and the single-particle Hamiltonian follows an ensemble that interpolates between the GOE and the Gaussian symplectic ensemble (GSE). The GSE describes grains with strong spin-orbit scattering, such as those made of gold~\cite{Bolotin2004,Kuemmeth2008}. The RMT ensembles are characterized by level repulsion (i.e., a suppressed probability to have two levels close to one another), which is stronger in the GSE than in the GOE. Level repulsion signifies the breaking of spatial symmetries, in which case  the single-particle levels are only two-fold 
degenerate  because of time-reversal symmetry. This two-fold degeneracy can be lifted by an external magnetic field. 

\subsection{Electron-electron interactions}

\subsubsection{Constant-interaction model}

The simplest model of electron-electron interaction is the constant-interaction (CI) model, in which the interaction is taken to be the charging energy $E_C \hat N^2$ where $\hat N$ is the particle-number operator and  $E_C=e^2/(2C)$ with $C$ the capacitance of the grain.

The positions (\ref{tunneling_spectr_energy}) of successive differential-conductance peaks are then given by a non-fluctuating $N$-dependent term plus single-particle energies. It follows that the fluctuations are completely determined by the single-particle spectrum and therefore satisfy RMT statistics. 

The CI model works well in gold, platinum, and copper~\cite{Petta2001,Kuemmeth2008}, but fails in superconducting or ferromagnetic grains. 

\subsubsection{Universal Hamiltonian}

The matrix elements of the electron-electron interaction in a chaotic grain fluctuate when expressed in the basis of the single-particle Hamiltonian. We can separate them into an average and fluctuating parts.  The matrix elements of the fluctuating part are typically smaller than the average interaction by a factor of $1/g_{\rm Th}$ where $g_{\rm Th} = E_{\rm Th} / \delta$. In the limit where $g_{\rm Th} \gg 1$, we can ignore the fluctuating part and the remaining average interaction together with the one-body Hamiltonian is the so-called universal Hamiltonian~\cite{Kurland2000, Aleiner2002}.

A simple way to derive the universal Hamiltonian is to argue that in a chaotic grain the interaction terms should be invariant under a change of the single-particle basis~\cite{Alhassid2005_prb}.  In the absence of spin-orbit coupling and orbital magnetic field, there are three interaction terms that are invariant under orthogonal transformations of the single-particle basis, leading to a universal Hamiltonian of the form
\begin{equation}\label{universal_Hamiltonian}
 \hat{H} =  \sum_{k, \sigma=\uparrow,\downarrow} \varepsilon_k c_{k\sigma}^\dagger
c_{k\sigma} + E_C \hat{N}^2 - G\hat{P}^\dagger P - J_s \hat{{\bf S}}^2 \;.
\end{equation}
Here $\varepsilon_k$ are spin-degenerate single-particle energies, $\bf S$ is the total spin of the electrons in the grain, and
\begin{equation}
 \hat{P}^\dagger = \sum_k c^\dagger_{k\uparrow}c^\dagger_{k\downarrow} \;
\end{equation}
is a pair creation operator in spin up-spin down orbitals.  

\begin{figure}
 \includegraphics[width=\columnwidth]{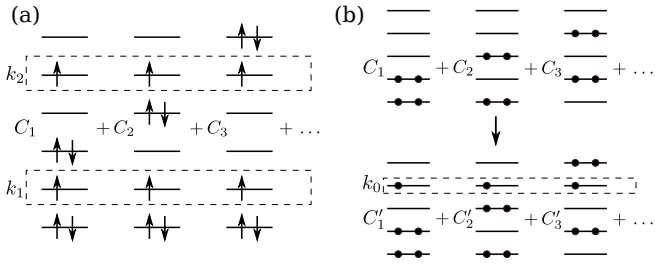}
 \caption{(a) An eigenstate of the universal Hamiltonian (\ref{universal_Hamiltonian}) in the absence of spin-orbit scattering with the total spin $S=1$. Each orbital can be occupied by one spin-up and one spin-down electrons. (b) Possible initial (top) and final (bottom) eigenstates of the pairing Hamiltonian for tunneling onto a grain with even particle number. In the presence of spin-orbit scattering, the single-particle eigenstates no longer have good spin projection, but they still come as Kramers doublets.}\label{Fig-states}
\end{figure}

Each eigenstate of the universal Hamiltonian (\ref{universal_Hamiltonian}) factorizes into a direct product of a fully paired state, i.e., a state in which singly occupied orbitals are excluded, and of a state with singly occupied orbitals only~\cite{Schmidt2007, Schmidt2008}
\begin{equation}\label{eigenstate_factorization}
|\mathcal{B},\zeta,\gamma,S,M\rangle = |\zeta\rangle_{\mathcal{U}}\otimes |\gamma,S,M\rangle_{\mathcal{B}} \;.
\end{equation}
Here $\mathcal{B}$ is the set of singly occupied orbitals, $\mathcal{U}$ is the set of the remaining (doubly occupied and empty) orbitals, $\zeta$ distinguishes states that are different by pair scattering, and $\gamma$ labels the different ways to couple the set $\mathcal{B}$ of spin-$1/2$ singly occupied levels to total spin $S$ and spin projection $M$.  An example of such a state with the total spin $S=1$ and spin projection $M=1$ is shown in Fig.~\ref{Fig-states}(a), where two singly occupied orbitals $k_1$ and $k_2$ are occupied by spin-up electrons. Note that the singly occupied orbitals are the same in each of the non-interacting states of the superposition. This is the so-called blocking effect of the pairing interaction~\cite{Soloviev1961}, which can only scatter a pair of electrons in a doubly occupied orbital to an empty orbital and therefore does not affect singly occupied levels. In contrast, only singly occupied orbitals contribute to the total spin and are thus relevant for the exchange 
interaction. The good-spin eigenstates are obtained by coupling $m$ spin-$1/2$ electrons in the set $\mathcal{B}$ of singly occupied levels to total spin $S$, and their corresponding energy eigenvalue is  $E_{\mathcal{B}} = \sum_{k\in \mathcal{B}}\varepsilon_k - J_s S(S+1)$.   The degeneracy of each such eigenvalue (not including the magnetic degeneragy of $2S+1$) is given by ${m \choose S+m/2} - {m \choose S+1+m/2}$, where ${m\choose n}$ is a combinatorial coefficient~\cite{Alhassid2003}. 

The fully paired eigenstates and their energies can be obtained by solving Richardson's equations in the subspace $\mathcal{U}$ of  doubly occupied and empty orbitals~\cite{Richardson1963,Richardson1967}. Such a state can be written as
\begin{equation}\label{eigenstate_pairing}
 |\zeta\rangle_{\mathcal{U}} \propto \prod_{\mu=1}^{m_c} \left(\sum_{k \in \mathcal{U}} \frac{1}{2\varepsilon_k - R_\mu} c^\dagger_{k\uparrow} c^\dagger_{k\downarrow}\right) |0\rangle\,,
\end{equation}
where $m_c$ is the number of Cooper pairs and $R_\mu$ are the Richardson parameters satisfying the following nonlinear equations:
\begin{equation}\label{Richardson_equation}
 \frac 1G + 2\sum_{\nu=1, \nu \ne \mu}^{m_c}\frac{1}{R_\nu - R_\mu} = \sum_{k \in \mathcal{U} } \frac{1}{2\varepsilon_k - R_\mu}\,.
\end{equation}
The energy contribution from such a state is given by $E_{\mathcal{U}} = \sum_{\mu=1}^{m_c}R_\mu$, so the total eigenenergy of the state (\ref{eigenstate_factorization}) is the sum  $E_{\mathcal{B}} + E_{\mathcal{U}} + E_C N^2$.

Spin-orbit coupling breaks spin-rotation symmetry but preserves time-reversal symmetry. Therefore, when we turn on a spin-orbit scattering term, the single-particle eigenstates are no longer eigenstates of the spin-projection operator but they still come in time-reversed degenerate pairs (Kramers degeneracy), which we denote by $k1$ and $k2$.  The one-body Hamiltonian makes a crossover from GOE to the GSE, and the exchange term, written in a basis diagonalizing one-body part, acquires a complicated structure~\cite{Gorokhov2003}. In the presence of sufficiently strong spin-orbit scattering,  the single-particle Hamiltonian follows GSE statistics.  The average interaction is now required to be invariant under symplectic transformations of the single-particle basis. This makes the spin-exchange term  incompatible with the symmetries of the one-body part and is therefore absent from the universal Hamiltonian. The pairing term is still compatible with the symmetries of the one-body Hamiltonian and remains 
unaffected.  The universal Hamiltonian now has the form
\begin{equation}\label{universal_Hamiltonian_soc}
 \hat{H} = \sum_{k\mu}\varepsilon_k c^\dagger_{k\mu}c_{k\mu}  + E_C \hat{N}^2 - G\hat{P}^\dagger P \,,
\end{equation}
where the index $\mu=1,2$ labels single-particle Kramers doublets, and
\begin{equation}
 \hat{P}^\dagger = \sum_k c^\dagger_{k2} c^\dagger_{k1} \;
\end{equation}
is a pair creation operator in the Kramers doublets.

 In this GSE limit, the blocking effect of the pairing interaction still holds. Similarly to Eq.~(\ref{eigenstate_factorization}), an eigenstate of (\ref{universal_Hamiltonian_soc}) factorizes into a fully paired state (which is a superposition of doubly occupied and empty orbitals), and a set of singly occupied levels (which can no longer be expressed as good-spin eigenstates). The contribution of the paired state to the energy is obtained by solving the corresponding Richardson's equations (\ref{Richardson_equation}) and the corresponding paired eigenstate is given by Eq.~(\ref{eigenstate_pairing}), in which we replace  $c^\dagger_{k\uparrow} c^\dagger_{k\downarrow}$ by $c^\dagger_{k1} c^\dagger_{k2}$. The contribution of the singly occupied levels to the energy eigenvalue  is simply given by the sum of their single-particle energies. 
  
 An example of a pair of  eigenstates participating in the tunneling process in the GSE limit of the universal Hamiltonian is shown in Fig.~\ref{Fig-states}(b). There, an electron tunnels onto a grain with an even particle number $N_e$ in a fully-paired state $|\Omega\rangle_{N_e}$. A possible final state has exactly one unpaired electron in a singly occupied orbital $k_0$.  Note that the initial and final states in a tunneling process may differ only by one singly occupied orbital.

\section{Magnetic response of energy levels}\label{Sec-response}

Here we consider the discrete energy spectrum of a grain in an external magnetic field $B$. Since $B$ couples to the electron's spin, the magnetic response of energy levels is sensitive to spin-orbit scattering. Such sensitivity was probed in several single-electron tunneling spectroscopy experiments~\cite{Salinas1999,Davidovic1999,Petta2001,Petta2002,Kuemmeth2008}.

\subsection{CI model: response of single-particle levels}

\begin{figure}
 \includegraphics[width=\columnwidth]{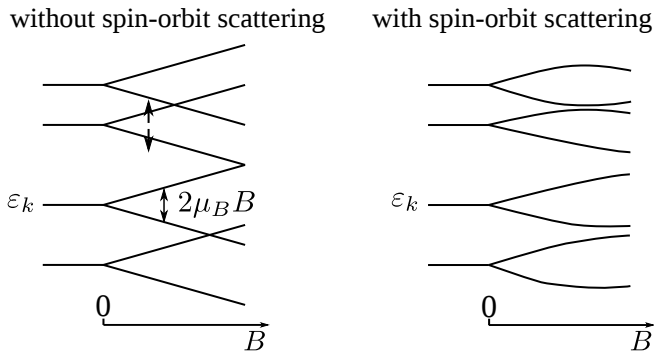}
 \caption{Splitting of single-electron Kramers doublets by an external magnetic field $B$ in the absence (left) and presence (right) of spin-orbit scattering. }\label{Fig-splitting}
\end{figure}

When a grain is well described by the CI model, the magnetic-field dependence of energies (\ref{tunneling_spectr_energy}) measured in a tunneling spectroscopy experiment is that of the single-particle levels. A finite Zeeman magnetic field $B$ lifts the Kramers degeneracy of the single-electron levels and splits their doublets as is shown schematically in Fig.~\ref{Fig-splitting}. 

 In the absence of spin-orbit scattering, spin is a good quantum number, and the levels split  as $\varepsilon_{k\pm}(B) = \varepsilon_k(0) \pm \frac 12 g\mu_B B$, where $g=2$ is the single-particle $g$ factor and $\mu_B$ is the Bohr magneton [see the left panel of Fig.~\ref{Fig-splitting}]. We note that orbital magnetism does not contribute to $g$ factors~\cite{Kravtsov1992,Matveev2000} and is usually ignored in nonlinear corrections to  $\varepsilon_{k\pm}(B)$. 
  
  In grains with spin-orbit scattering, the $g$ factor fluctuates from level to level~\cite{Matveev2000,Brouwer2000,Adam2002}, depends on the direction of $B$~\cite{Brouwer2000,Adam2002}, and, in general, acquires a finite orbital contribution~\cite{Matveev2000,Adam2002,Cehovin2004}. Its statistics were calculated using the RMT statistics of the single-particle wavefunctions~\cite{Matveev2000, Brouwer2000, Adam2002} \, and were found to agree with measurements in noble-metal nanoparticles~\cite{Petta2001,Petta2002,Kuemmeth2008}. In the GSE limit, the $g$-factor distribution is given by~\cite{Matveev2000}
  \begin{equation}\label{gfact-distr-nonint}
   P(g) = 3\sqrt{\frac{6}{\pi}} \frac{g^2}{\left\langle g^2\right\rangle^{3/2}}\exp\left(-\frac{3g^2}{2\left\langle g^2\right\rangle}\right) \;.
  \end{equation}
  
 In addition, the spin operator now couples different Kramers doublets $k$, which leads to nonlinear corrections to
 $\varepsilon_k$ and to avoided crossing of energy levels [see the right panel of Fig.~\ref{Fig-splitting}]. Up to quadratic corrections, which are parametrized by the level curvature $\kappa_k$, we write 
\begin{equation}\label{definition_sp}
 \varepsilon_{k\pm}(B) = \varepsilon_k(0) \pm \frac 12 g\mu_B B + \frac 12 \kappa_k B^2 + O(B^3)\,.
\end{equation}
The distribution of level curvatures $\kappa_k$ can be calculated from RMT~\cite{Fyodorov1995, vonOppen1995, Mucciolo2006}, and in the GSE limit
\begin{equation}\label{curv-distr-nonint}
 P(\kappa) = \frac{8}{3\pi\sqrt{3\langle\kappa^2\rangle}} \frac{1}{\left[1+{\kappa^2}/\left({3\langle\kappa^2\rangle}\right)\right]^3} \;.
\end{equation}
This distribution is shown by the solid line in Fig.~\ref{Fig-1st-peak} and agrees with the experimental results of Ref.~\cite{Kuemmeth2008} for gold nanoparticles. We note that RMT does not predict the average values $\left\langle g^2\right\rangle$ and $\langle \kappa^2\rangle$ in these distributions, but they are related by~\cite{Nesterov2015}
\begin{equation}\label{kappa-g-relation}
 \frac{\sqrt{\langle \kappa^2 \rangle}}{\mu_B^2/\delta} = \frac{\pi }{3\sqrt{3}} \langle g^2\rangle \;.
\end{equation}

\subsection{Effects of electron-electron interactions}

\subsubsection{Many-particle g factor and level curvature}

In the presence of electron-electron interactions beyond the charging energy of the CI model, the energy differences in (\ref{tunneling_spectr_energy}) do not reduce to single-particle quantities.  The single-particle definition  (\ref{definition_sp}) of $g$ factors and level curvatures has to be generalized to the many-particle levels. Here we discuss the case of one-bottleneck geometry, in which the barrier 
between the source electrode and the grain is much thicker than the second barrier, and the grain has sufficient time to relax to its ground state before the next tunneling event~\cite{vonDelft2001, Gorokhov2003, Gorokhov2004}. If electron tunneling occurs onto a grain with an even electron number $N_e$, one measures an energy difference $\Delta E_{\Omega,0} = E^{N_e + 1}_{\Omega'} - E^{N_e}_0$ between the energy of a twofold-degenerate state $|\Omega'\rangle_{N_e+1}$ of the odd particle number $N_e+1$ and the energy of the non-degenerate even ground state $|0\rangle_{N_e}$. We then define the many-particle $g$ factor and level curvature $\kappa$ by
\begin{equation}\label{definition_mb}
 \Delta E_{\Omega,0}(B) = \Delta E_{\Omega,0}(0) \pm \frac 12 g \mu_B B + \frac{1}{2} \kappa B^2 + O(B^3)\,.
\end{equation}

For the CI model, expression (\ref{definition_mb}) reduces to the single-particle level expression (\ref{definition_sp}). 

\subsubsection{Exchange interaction}

The effect of the exchange interaction on the statistics of the
$g$ factor [defined according to Eq.~(\ref{definition_mb})]  was studied in Refs.~\cite{Gorokhov2003, Gorokhov2004} in the crossover between GOE and GSE and in the absence of pairing correlations.
A many-body eigenstate is generally a complex superposition of states with different occupation numbers,  and the $g$-factor distribution differs from its single-particle distribution in the crossover. 
However, in the GSE limit, the effective exchange interaction is suppressed, and the $g$-factor distribution reduces to the its single-particle GSE distribution. This is consistent with incompatibility of the exchange term with the symmetries of the one-body part of the universal Hamiltonian in the GSE limit.  Large $g$ factors in ferromagnetic Co nanoparticles were recently measured in Ref.~\cite{Gartland2013}.

\subsubsection{Pairing interaction}

The effects of the pairing interaction on both the $g$ factor and level curvature statistics were studied in Ref.~\cite{Nesterov2015}. It was shown that the many-particle $g$ factor reduces to its value for the blocked level, and thus its distribution is not modified by pairing correlations. However, it was found that the level curvature statistics are very sensitive to pairing correlations. In the following we briefly review the main ideas and results of Ref.~\cite{Nesterov2015}.

We first discuss the $g$ factor. Its independence of pairing correlations is a direct consequence of time-reversal symmetry and of the blocking effect of pairing. The ground state of a grain with an even number of electrons in the presence of pairing correlations is a superposition of Slater determinants in which the single-particle levels are either doubly occupied or empty. When an electron tunnels onto the grain in such a state, a possible final state contains one unpaired electron in one blocked orbital [see, e.g., Fig.~\ref{Fig-states}(b)].  The magnetic moment of the fully paired even ground state is exactly zero since under time reversal the even ground state is even while the magnetization operator is odd. In addition, fully paired electrons in the odd state do not contribute to its magnetic moment. Therefore,
 the magnetic moment in the odd state reduces to the single-particle magnetic moment of the electron in the blocked orbital and so does the $g$ factor. We note that the above conclusion follows from the general structure of the pairing interaction. It is independent of the statistics of the single-electron states and of the particular form of the magnetic-moment operator (i.e., of the orbital contribution to the magnetic moment). 

We next turn to level curvatures. In the absence of pairing correlations, $\kappa$ is the single-particle level curvature, for which second-order perturbation theory gives
\begin{equation}\label{kappa-sp}
 \kappa_{k_0} = 2\sum_{k\ne k_0} \frac{|M^z_{k_01,k1}|^2 + |M^z_{k_01, k2}|^2}{\varepsilon_{k_0} - \varepsilon_k} \;.
\end{equation}
Here, $k_0$ is the singly occupied orbital in the final state,  $M^z_{k\alpha,k'\alpha'}$ is the single-particle matrix element of the magnetic-moment operator between the states $|k\alpha\rangle$ and $|k'\alpha'\rangle$, and the sum runs over the pairs of single-particle states of different doublets. On average, such curvature is zero, and its distribution is symmetric. In the GSE limit, this distribution is given by Eq.~(\ref{curv-distr-nonint}) and it has power-law tails $P(\kappa_0) \sim 1/\kappa^6_{k_0}$~\cite{vonOppen1995} because of the power law for level repulsion. 

In the presence of pairing correlations, the level curvature for the transition $|0\rangle_{N_e} \rightarrow |\Omega\rangle_{N_e+1}$ is given by the second-order corrections to the two eigenenergies
\begin{equation}\label{kappa-difference}
 \kappa = \kappa_\Omega^{N_e+1} - \kappa_0^{N_e}\,,
\end{equation}
where
\begin{equation}\label{kappa-odd}
 \kappa_\Omega^{N_e+1} = 2\sum'_{\Omega'} \frac{|\langle \Omega|\hat{M}_z|\Omega'\rangle_{N_e+1}|^2}{E_\Omega^{N_e+1} - E_{\Omega'}^{N_e+1}}
\end{equation}
and 
\begin{equation}\label{kappa-even}
 \kappa_0^{N_e} = 2\sum'_{\Theta'} \frac{|\langle 0|\hat{M}_z|\Theta'\rangle_{N_e}|^2}{E_{0}^{N_e} - E_{\Theta'}^{N_e}}\,.
\end{equation}
The sums in (\ref{kappa-odd}) and (\ref{kappa-even}) run over the states with energies different than $E_\Omega^{N_e+1}$ and $E_{0}^{N_e}$, respectively.  The quantities defined in these equations can be thought of as the curvatures of the odd and even states. 

When the final state $|\Omega\rangle_{N_e+1}$ is the odd ground state $|0\rangle_{N_e+1}$, both contributions (\ref{kappa-odd}) and (\ref{kappa-even}) are negative since they arise from second-order corrections to the ground-state energies. In the non-interacting limit, their difference reduces to Eq.~(\ref{kappa-sp}), so their average values are equal in the absence of pairing.  In the presence of strong pairing with $\Delta/\delta \gg 1$, there is a gap in the even excitation spectrum, which suppresses the even contribution (\ref{kappa-even}). However, no such gap exists in the odd spectrum, so the odd contribution (\ref{kappa-odd}) is not suppressed by pairing. Moreover,  the lowest excitation energy for an odd particle number can be estimated as $\sqrt{\delta^2 +\Delta^2} - \Delta \sim \delta^2/(2\Delta) \ll \delta$. This excitation energy decreases in the presence of pairing and enhances the odd contribution.  Therefore, for the transition between two ground states, pairing correlations  enhance the 
negative contributions and suppress the positive 
contributions in Eq.~(\ref{kappa-difference}), leading to an asymmetric distribution with an overall negative average.  For a fluctuating single-particle spectrum, the enhancement of the odd contribution makes the curvature fluctuations stronger, and thus makes the left tail of the level curvature distribution longer. 

\begin{figure}
 \includegraphics[width=\columnwidth]{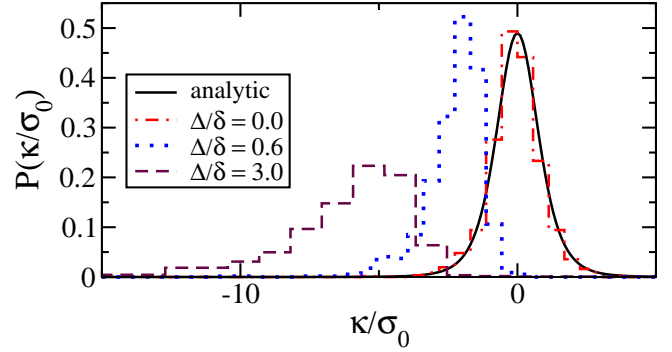}
 \caption{Histograms: the level-curvature distribution for a transition between two ground states (the first differential-conductance peak) calculated   for three different values of $\Delta/\delta$~\cite{Nesterov2015}. Solid line: analytic distribution (\ref{curv-distr-nonint}) of single-particle level curvatures in the GSE limit~\cite{vonOppen1995, Fyodorov1995}.  The curvature $\kappa$ is expressed in units of $\sigma_0$, the standard deviation of the single-particle level curvature. Adapted from Ref.~\cite{Nesterov2015}.}\label{Fig-1st-peak}
\end{figure}

These qualitative observations, which are generally valid in the $\Delta/\delta \gg 1$ limit, are well justified both by exact numerical calculations and by calculations based on a BCS-like approach~\cite{Nesterov2015}. In those simulations, the GSE limit was assumed and orbital contribution to the magnetic moment was ignored.  In Fig.~\ref{Fig-1st-peak}, we show the level-curvature distributions calculated from the exact numerical simulations for the transition between two ground states, which normally corresponds to the first differential-conductance peak. The effect of pairing correlations on the level-curvature distribution is already visible in the fluctuation-dominated regime of $\Delta/\delta <1$ (see the results for $\Delta/\delta = 0.6$). 

\begin{figure}
 \includegraphics[width=\columnwidth]{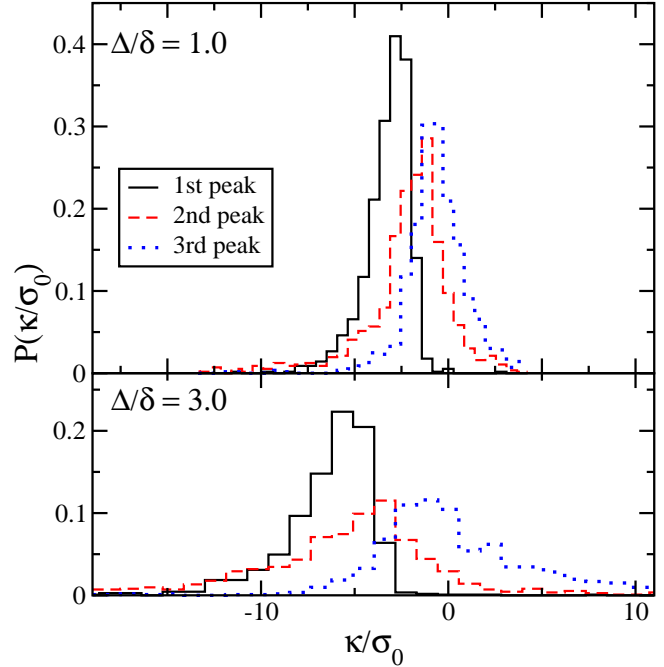}
 \caption{The level-curvature distributions $P(\kappa/\sigma_0)$ for the first three resolvable differential-conductance peaks for $\Delta/\delta = 1.0$ (top panel) and $\Delta/\delta = 3.0$ (bottom panel).  }\label{Fig-higher-peaks}
\end{figure}

It is also interesting to determine the statistics of tunneling into excited states of an odd grain. To this end, we first discuss which odd states are measurable in an experiment. In the non-interacting limit, an electron can only tunnel into an empty orbital, which lies above the Fermi level [see Fig.~\ref{Fig-intro}(b)]. In the presence of pairing correlations, electrons can also tunnel into orbitals below the Fermi level as long as they are within an energy window $\sim \Delta$. The height of the corresponding differential-conductance peak decreases when the energy of the blocked orbital of the final state moves further down, and the peak eventually becomes unresolvable in the experiments (see Ref.~\cite{Nesterov2015} for the calculation of the corresponding peak heights). In the numerical simulations, the resolvability threshold was taken to be 10\% of the average peak height of the non-interacting limit. In Fig.~\ref{Fig-higher-peaks} we show the corresponding level-curvature distributions of the first 
three resolvable peaks.

When $\Delta/\delta$ is sufficiently large, the second and third peaks typically correspond to the transitions into the states with the blocked orbitals $k_F-1$ and $k_F +1$, where $k_F$ is the blocked orbital for the first peak (odd particle number Fermi level). In the BCS limit, both of them and the odd ground state can be described as one-quasiparticle excitations on top of the even ground state with excitation energies $\sqrt{(\varepsilon_k -\mu)^2 +\Delta^2}$.  Therefore,  the excitation energies of these two states taken with respect to the odd ground-state energy can be estimated as $\sqrt{(\varepsilon_{k_F\pm 1} - \varepsilon_{k_F})^2 + \Delta^2} - \Delta \sim \delta^2/(2\Delta)$. We observe that the energies of these two states, $E^{N_e+1}_2$ and $E^{N_e+1}_3$, tend to be closer to one another than to energies of other odd states contributing to the odd curvature (\ref{kappa-odd}). The contribution from this pair of states has the same amplitude but opposite signs for the second and third peaks. 
Thus, one finds that the distribution of curvatures for the third peak is shifted to the right, and the positive tail is enhanced in comparison to that for the second peak. The dispersions of the curvatures of the second and third peaks are both larger than the dispersion for the first peak. When  the peak number increases further, the distribution approaches the single-particle (non-interacting) distribution, which corresponds to an electron tunneling into orbitals far above the Fermi level. 

In conclusion, $g$-factor statistics can be used as a tool to probe whether pairing interactions are sufficient to describe electron correlations in a  metallic grain, while level-curvature statistics can be used to detect whether pairing interactions are present at all. We emphasize that level curvatures are sensitive to pairing correlations even in the fluctuation-dominated regime $\Delta < \delta$, where a gap cannot be detected in the energy spectrum of the even-particle grain. The advantage of level curvatures is that they can be directly measured in the single-electron tunneling spectroscopy experiments. This is in contrast to thermodynamic observables~\cite{Falci2000, DiLorenzo2000, VanHoucke2006, Nesterov2013}, which are difficult to measure in a single grain. 

\section{Spin susceptibility}\label{Sec-thermodyn}

\begin{figure}
 \includegraphics[width=\columnwidth]{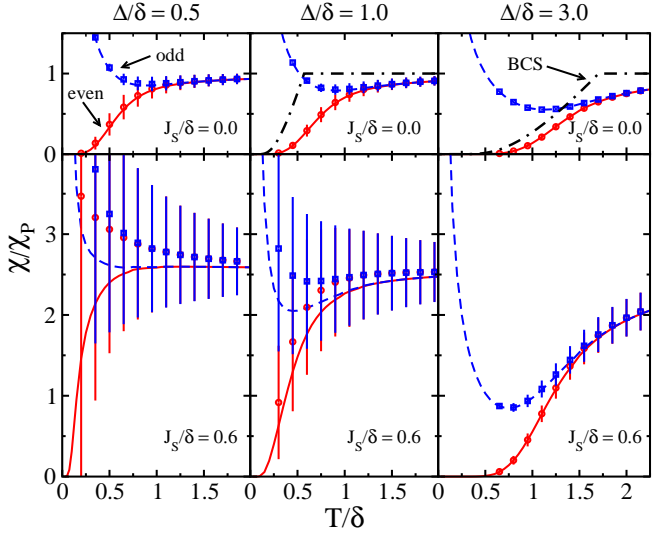}\caption{Spin susceptibility $\chi/\chi_P$ of a grain without spin-orbit scattering as a function of temperature $T/\delta$ for different values of $\Delta/\delta$ and $J_s/\delta$ and for an even and odd particle numbers. The symbols with error bars show the results for fluctuating RMT spectrum, while the solid and dashed lines correspond to an equally spaced single-particle spectrum. The calculations are based on the SPA+RPA method (see text), except for the low-temperature results for an equally spaced which are calculated using Richardson's solution (see Sec.~\ref{Sec-model}).  The dot-dashed lines are grand canonical BCS results for an equally spaced spectrum. Here $\chi_P = 2\mu_B^2/\delta$ is the Pauli susceptibility. Adapted from Ref.~\cite{Nesterov2013}.}\label{Fig-susc}
\end{figure}

In this section we consider the spin susceptibility, a thermodynamic observable that is particularly sensitive to the electron's spin and, therefore, to the presence of spin-orbit scattering in the grain. It is defined as the derivative of the magnetization with respect to external magnetic field (evaluated at zero field).  At inverse temperature $\beta=1/T$ (taking the Boltzmann constant to be $k_B = 1$), it is given by
\begin{equation}\label{susc-definition}
 \chi =   4 \mu_B^2 \int \limits_0^\beta d\tau \left\langle \hat{S}_z(\tau)\hat{S}_z(0)\right\rangle \,,
\end{equation}
where $\left\langle \hat{S}_z(\tau)\hat{S}_z(0)\right\rangle$ is the imaginary-time response function of $\hat S_z$  [here $\hat{S}_z(\tau) = e^{\tau\hat{H}}\hat{S}_z e^{-\tau\hat{H}}$, where $\hat{H}$ is the grain's Hamiltonian without magnetic field].  Its spectral representation in terms of the  many-body eigenstates $m$ and eigenvalues $E_m$ of $\hat{H}$ is
\begin{multline}\label{susc-general}
 \chi =  4\mu_B^2 \left(\beta \sum_m e^{-\beta E_m} \frac{\langle m | \hat{S}_z | m \rangle^2}{Z} \right.  \\ \left. +  \sum_{m\neq n} \frac{e^{-\beta E_m} - e^{-\beta E_n}}{E_n - E_m} \frac{\left|\langle n|\hat{S}_z|m\rangle\right|^2}{Z}\right)\,,
\end{multline}
where $Z=\sum_m e^{-\beta E_m}$ is the partition function.

\subsection{Grains without spin-orbit scattering}

When spin is a good quantum number, only the first term in Eq.~(\ref{susc-general}) contributes and it is given by $ 4 \mu_B^2  \beta \langle \hat S_z^2\rangle$. At low temperatures, $\chi$ is then determined by the ground-state spin and by the excitation energies of low-lying states with other spin values, leading to an odd-even effect in the number parity of electrons. In particular, for an even particle number and not too strong exchange interaction, there is an exponential suppression $\chi_e \sim \exp(- E_1/T)$, where $E_1$ is the lowest excitation energy of the $S=1$ triplet states, while for an odd particle number the spin susceptibility displays a Curie-like divergence $\chi_o \sim 1/T$. Since pairing correlations increase the excitation energies of states with higher spin values (describing broken Cooper pairs), they generally suppress $\chi$. Therefore, for even particle number and for large $\Delta/\delta$, one finds $E_1 \sim \Delta$ and $\chi_e \sim \exp(-\Delta/T)$ at low $T$. For odd particle 
number, the interplay between the $1/T$ 
divergence and the above suppression results in a re-entrant behavior of $\chi$ (i.e., a local minimum vs. temperature). The latter behavior was proposed as a signature of pairing correlations, which exists even in the fluctuation-dominated regime $\Delta/\delta <1$~\cite{DiLorenzo2000,Falci2000}. 

The spin susceptibility for a grain described by the universal Hamiltonian (\ref{universal_Hamiltonian}) was calculated in Ref.~\cite{Nesterov2013}. There, the exchange interaction was treated exactly using a spin-projection technique~\cite{Alhassid2003,Alhassid2007_prl}, and the pairing interaction was taken into account using the Hubbard-Stratonovich representation in terms of fluctuating pairing fields~\cite{Hubbard1959,Stratonovich1957}.  The spin-projection technique relies on the following identity, valid for a scalar operator $\hat{X}$
\begin{equation}\label{spin-projection-identity}
 {\rm Tr}_S \hat{X} = {\rm Tr}_{S_z=S} \hat{X} - {\rm Tr}_{S_z=S+1} \hat{X} \;.
\end{equation}
Here, ${\rm Tr}_S$ denotes the trace over states with a fixed total spin $S$, while ${\rm Tr}_{S_z = M}$ denotes the trace over states with a fixed spin component $S_z = M$. Using (\ref{spin-projection-identity}) we can write the partition function and spin susceptibility in terms of the $S_z$-projected partition functions $ {\rm Tr}_{S_z=M} e^{-\beta \hat{H}_{\rm BCS}}$ for the reduced Hamiltonian $\hat{H}_{\rm BCS}$ obtained from (\ref{universal_Hamiltonian}) when the exchange term is omitted. The propagator for this reduced BCS Hamiltonian is then written as a functional integral over the complex $\tau$-dependent pairing field $\widetilde{\Delta}$ as
\begin{multline}
 \exp \left[-\beta\left(\hat{H}_{\mathrm{BCS}} - \mu\hat{N}\right)\right] \\
 = \int \mathcal{D}[\widetilde{\Delta},\widetilde{\Delta}^*] \mathcal{T} \exp \left[- \int\limits_0^\beta d\tau \left( \frac{|\widetilde{\Delta}(\tau)|^2}{G} + \hat{H}_{\widetilde{\Delta}(\tau)}\right)\right]\,.
\end{multline}
Here
\begin{multline}
 \hat{H}_{\widetilde{\Delta}} = \sum_k \left[ \left(\epsilon_k - \mu- \frac G2\right)(c^\dagger_{k\downarrow} c_{k\downarrow} + c^\dagger_{k\uparrow}c_{k\uparrow}) \right. \\ \left. - \widetilde{\Delta} c^\dagger_{k\uparrow} c^\dagger_{k\downarrow} - \widetilde{\Delta}^* c_{k\downarrow} c_{k\uparrow} + \frac G2\right]
\end{multline}
describes non-interacting electrons in an external pairing field and $\mu$ is the chemical potential. 

The integral over static  ($\tau$-independent) values of the pairing field $\widetilde{\Delta}$ was evaluated exactly (known as the static-path approximation or SPA~\cite{Muhlschlegel1972,Alhassid1984,Lauritzen1988}), and the integral over small-amplitude time-dependent fluctuations of the pairing field was evaluated in the saddle-point approximation around each static value of the pairing field. This approximation is known as the SPA plus random-phase approximation (RPA)~\cite{Kerman1981,Kerman1983,Puddu1991,Lauritzen1993,Rossignoli1997,Attias1997}. To account for  odd-even effects, a number-parity projection was used~\cite{Goodman1981,Rossignoli1998,Balian1999,Alhassid2005_prc}.

The SPA plus RPA  breaks down at the lowest temperatures, when a certain RPA correction becomes unstable. When the approximation is stable, it is found to be quite accurate when compared with exact simulations~\cite{Nesterov2013}.

The statistical fluctuations of the spin susceptibility, calculated in the SPA+RPA approximation, are demonstrated in Fig.~\ref{Fig-susc}. In addition to the already mentioned number-parity signatures of pairing correlations, we notice the effects of exchange interaction, which generally competes with the pairing term. This competition is especially prominent in the fluctuation-dominated regime, where exchange correlations lead to much stronger mesoscopic fluctuations and to a suppressed odd-even effect. At higher temperatures, however, the exchange interaction may enhance the re-entrant behavior of the odd spin susceptibility. It increases $\chi$ in the high-$T$ limit and therefore increases the visibility of the local minimum.

\subsection{Grains with spin-orbit scattering}

We now discuss how the signatures of pairing correlations in the spin susceptibility are modified in grains with spin-orbit scattering. In the presence of spin-orbit coupling, spin-rotation symmetry is broken, and the full expression (\ref{susc-general}) must be used. As a result, the previous arguments in favor of the exponential suppression of $\chi_e$ at $T=0$ in grains with zero ground-state spin are no longer valid. In addition, the Curie-like divergence at low $T$ is now reduced because of the smaller contribution from the first term on the r.h.s. of (\ref{susc-general}). Therefore, the odd-even effect in $\chi$ is generally expected to be suppressed by finite spin-orbit scattering. Spin-orbit scattering also reduces the effects of exchange, namely, the strong fluctuations of $\chi$ at $\Delta <\delta$ and the enhanced re-entrant behavior at $\Delta > \delta$. 

The spin susceptibility of superconductors with spin-orbit scattering at $T=0$ was studied theoretically in the context of the anomalous Knight shift~\cite{Ferrell1959, Anderson1959_prl}. There, it was demonstrated that spin-reversing scattering can explain the deviation from the grand-canonical BCS theory predictions of vanishing susceptibility. In particular, it was demonstrated that $\chi$ can assume values close to those of normal-metal samples. In ultrasmall grains, it would imply $\chi \sim \chi_P$, where $\chi_P = 2\mu_B^2/\delta$ is the Pauli susceptibility.

The temperature dependence of $\chi$ in small superconducting and normal-metal particles with spin-orbit scattering was studied in Ref.~\cite{Shiba1976}, using an equally spaced single-particle spectrum. The calculations were based on the SPA in the grand-canonical  ensemble. It was shown that, as the strength of the spin-orbit scattering increases, $\chi(T)$ changes from being exponentially suppressed to being almost temperature-independent $\chi(T) \approx \chi_P$.

Finally, the odd-even effect in the spin susceptibility of small normal-metal particles with an equally spaced single-particle spectrum was studied in Ref.~\cite{Sone1977}. There, the behavior of $\chi$ for an even number of electrons resembled the grand-canonical results of Ref.~\cite{Shiba1976}. For odd particle number, it was shown that the $1/T$ divergence becomes suppressed by spin-orbit scattering since the expectation value of the ground-state spin decreases, and $\chi$ eventually approaches $\chi_P$ for strong spin-orbit coupling. Therefore, in the limit of strong spin-orbit scattering, $\chi \approx \chi_P$ for both even and odd grains at all temperatures, and the odd-even effect gets washed out. 

 We note that,  for $T\rightarrow 0$ and with negligible orbital magnetism, the second contribution in Eq.~(\ref{susc-general}) gives the odd or even ground-state level curvature up to a constant factor [see Eqs.~(\ref{kappa-odd}) and (\ref{kappa-even})]. When the first contribution in (\ref{susc-general}) is suppressed by strong spin-orbit scattering, the difference between odd and even spin susceptibilities is thus proportional to the level curvature (\ref{kappa-difference}) measured in a spectroscopy experiment for the first conductance peak. 

\section{Conclusions}\label{Sec-conclusions}

We have reviewed the interplay between pairing correlations and spin-orbit scattering in nanoscale metallic grains. Of particular interest is the crossover between the bulk BCS and the fluctuation-dominated regimes. This crossover can be studied experimentally by using grains of different sizes, thus modifying the ratio $\Delta/\delta$.

The magnetic response of discrete energy levels provides an observable to test both whether the pairing interaction is sufficient to describe nontrivial components of electron-electron correlations, and whether pairing correlations are present at all. The $g$ factors are independent of pairing correlations but get modified by, for example, an exchange interaction, and therefore can be used to test the importance of interactions other than pairing. Since level curvatures are very sensitive to pairing correlations, they can be used to probe them. 

On the other hand, signatures of pairing correlations in the spin susceptibility get suppressed by finite spin-orbit scattering and might be washed out when spin-orbit scattering is strong. Thus the spin susceptibility might not be a good signature of pairing correlations in the presence of spin-orbit scattering. Nevertheless, it would be interesting to calculate explicitly the spin susceptibility and its mesoscopic fluctuations in nanoscale superconducting grains using methods similar to the methods we used in the absence of spin-orbit scattering.

\begin{acknowledgements}
This work was supported in part by the U.S. DOE grant No. DE-FG02-91ER40608. Computational cycles were provided by the Yale University Faculty of Arts and Sciences High Performance Computing Center.
\end{acknowledgements}


\providecommand{\WileyBibTextsc}{}
\let\textsc\WileyBibTextsc
\providecommand{\othercit}{}
\providecommand{\jr}[1]{#1}
\providecommand{\etal}{~et~al.}

\end{document}